\documentclass{appolb}
\usepackage{epsfig}
\usepackage[auth-sc]{authblk}

\begin{document}
\title{ The ArDM experiment %
\thanks{Presented at the Cracow Epiphany Conference on Physics in Underground Laboratories and Its Connection with LHC. }%
}

\author[h]{M.Hara\'{n}czyk}
\author[b]{C.Amsler}
\author[a]{A.Badertscher}
\author[b]{V.Boccone}
\author[k]{N.Bourgeois}
\author[c]{A.Bueno}
\author[c]{M.C.Carmona-Benitez}
\author[g]{M.Chorowski}
\author[b]{W.Creus}
\author[a]{A.Curioni}
\author[f]{E.Daw}
\author[a]{U.Degunda}
\author[b]{A.Dell'Antone}
\author[a]{M.Dr\"{o}ge}
\author[a]{L.Epprecht}
\author[a]{C.Haller}
\author[a]{S.Horikawa}
\author[a]{L.Kaufmann}
\author[i]{J.Kisiel}
\author[a]{L.Knecht}
\author[a]{M.Laffranchi}
\author[e]{J.Lagoda}
\author[a]{C.Lazzaro}
\author[f]{P.Lightfoot}
\author[c]{J.Lozano}
\author[a]{D.Lussi}
\author[k]{G.Maire}
\author[i]{S.Mania}
\author[a]{A.Marchionni}
\author[j]{K.Mavrokoridis}
\author[c]{A.Melgarejo}
\author[e]{P.Mijakowski}
\author[a]{G.Natterer}
\author[c]{S.Navas-Concha}
\author[b]{P.Otiougova}
\author[g]{A.Piotrowska}
\author[g]{J.Poli\'{n}ski}
\author[d]{M.de Prado}
\author[e]{P.Przew\l ocki}
\author[k]{S.Ravat}
\author[b]{C.Regenfus}
\author[a]{F.Resnati}
\author[f]{M.Robinson}
\author[b]{J.Rochet}
\author[d]{L.Romero }
\author[e]{E.Rondio}
\author[a]{A.Rubbia}
\author[b]{L.Scotto-Lavina}
\author[f]{N.Spooner}
\author[a]{T.Viant}
\author[e]{A.Trawi\'{n}ski}
\author[a]{J.Ulbricht}
\author[h]{A.Zalewska}
\affil[a]{ETH Zurich, Switzerland}
\affil[b]{ Zurich University, Switzerland}
\affil[c]{ University of Granada, Spain}
\affil[d]{CIEMAT, Spain}
\affil[e]{A. Soltan Institute for Nuclear Studies, Warszawa, Poland}
\affil[f]{University of Sheffield, United Kingdom}
\affil[g]{ Wroclaw University of Technology, Wroclaw, Poland}
\affil[h]{H. Niewodnicza\'{n}ski Institute of Nuclear Physics PAN, Krakow, Poland}
\affil[i]{University of Silesia, Katowice, Poland}
\affil[j]{University of Liverpool, United Kingdom}
\affil[k]{ CERN, Switzerland  }

\maketitle
\begin{abstract}
The aim of the ArDM project is the development and operation of a one ton double-phase liquid argon detector for direct Dark Matter searches. The detector measures both the scintillation light and the ionization charge from ionizing radiation using two independent readout systems. This paper briefly describes the detector concept and presents preliminary results from the ArDM R\&D program, including a 3 l prototype developed to test the charge readout system.

\end{abstract}
\PACS{ 95.35.+d }

\section{Introduction}\label{secIntroduction}

Understanding the nature of Dark Matter is a key issue in contemporary particle physics and astrophysics. One of the prime candidates are the Weakly Interacting Massive Particles - WIMPs.
Most of the Dark Matter models predict that WIMPs interact only
gravitationally and weakly with matter, thus forming cold relic
halos around massive astronomical objects. Though there is
convincing evidence for Dark Matter from cosmology and astronomy,
WIMP interactions on ordinary matter are exceedingly rare and
therefore direct detection of Dark Matter particles is very
difficult and to-date no uncontroversial signal has been
discovered \cite{Mambrini},\cite{Szelc}. From the experimental point of view, noble liquids, such as argon and xenon, are one of the best options for large WIMP direct detection experiments. They combine the possibility of reaching ton-scale target masses, low energy threshold (in the keV), and the necessary discrimination power against background. The Argon Dark Matter experiment (ArDM) was initiated
in 2004 and is being developed at CERN. The goal of the project is
to assemble and operate the 1 ton scale double phase (liquid-gas)
argon detector with independent readout of the primary
scintillation light and the ionization charge. The principle of
design and operation of the ArDM detector is described in section
\ref{secArDMdesign}.

\section{Design and first test of the ArDM detector}\label{secArDMdesign}
\subsection{Detector description }
\begin{figure}[htp]
\centering
\includegraphics[angle=90,scale=0.26]{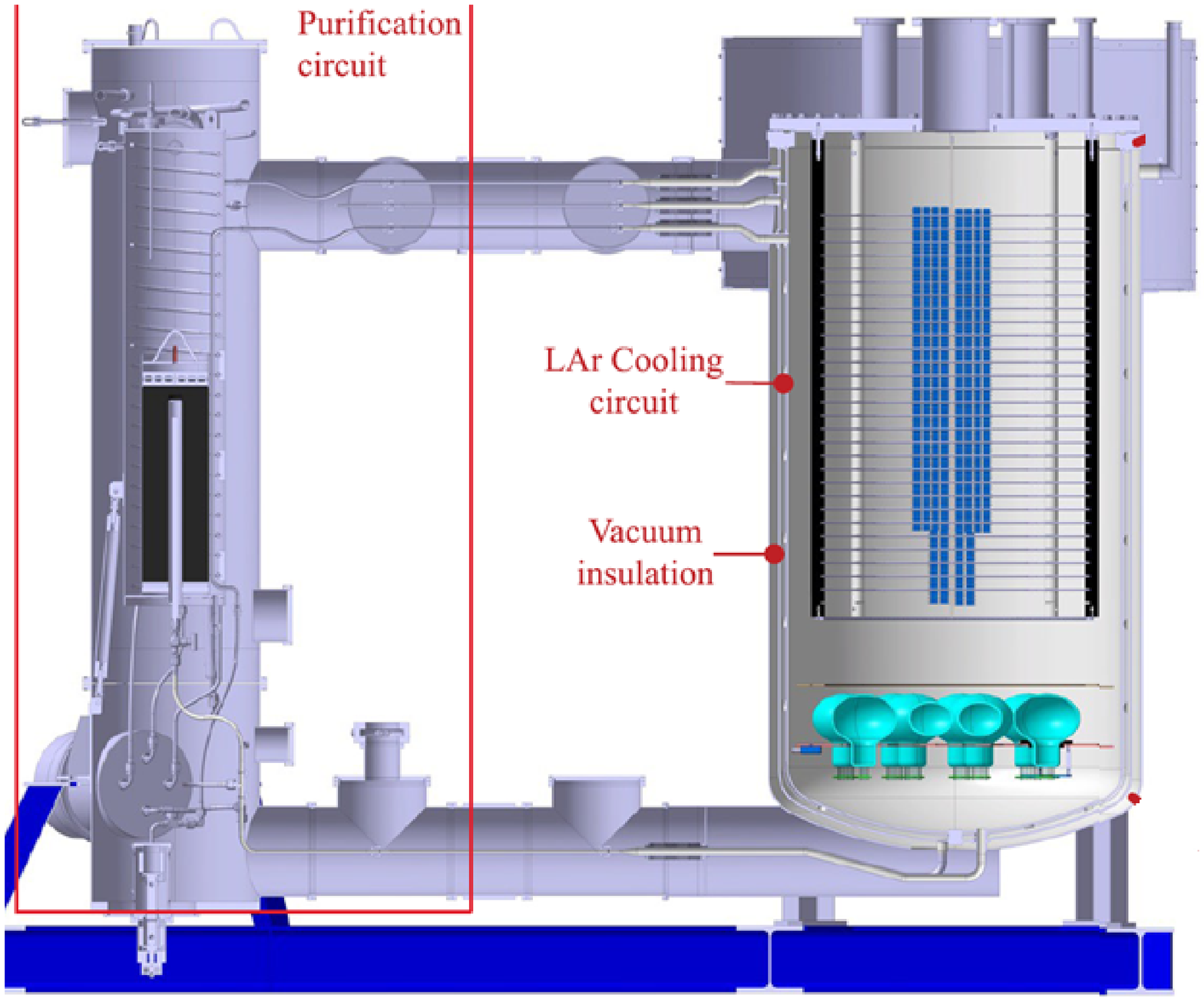}
\includegraphics[angle=0,scale=0.16]{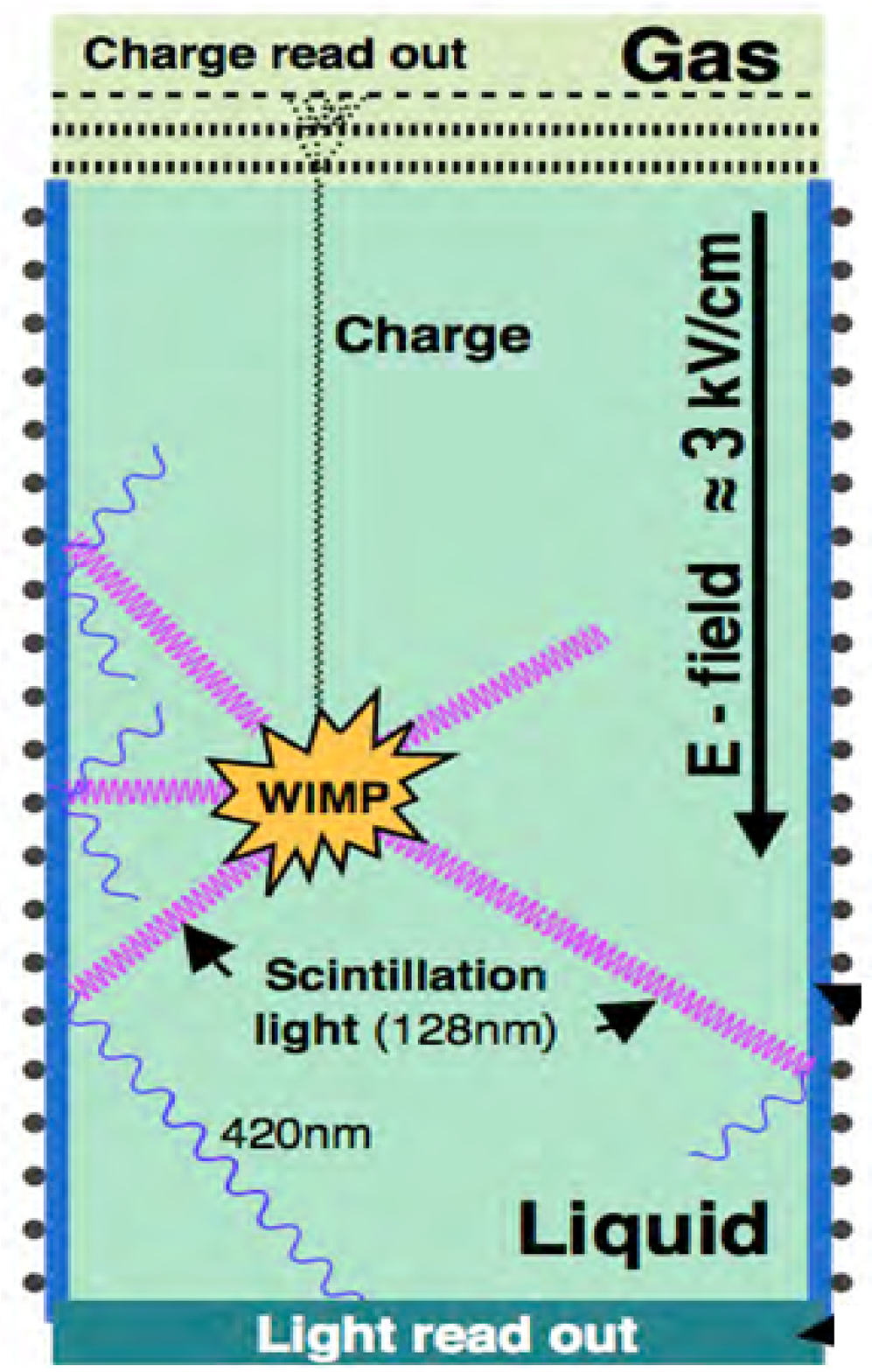}
\caption{\label{Fig.design} Design and working principle of the double-phase Argon Dark Matter detector.}
\end{figure}

The conceptual design of the ArDM experiment is illustrated in
Fig.\ref{Fig.design}. A vacuum insulated, stainless steel dewar,
200 cm high and 100 cm in diameter, contains a sensitive mass of
850 kg of liquid argon. In order to create a strong uniform
electric field along a vertical drift of the length of 120 cm, a
Greinacher (Cockroft-Walton) chain keeps the cathode, placed at
the bottom of the sensitive region, and the field shaping rings at
the proper voltage. Fourteen 8'' low background Photo Multiplier Tubes (PMT), suitable for cryogenic operation, are used for light readout. They are placed below the cathode grid. In liquid argon, the scintillation light is
produced in a narrow bandwidth around 128 nm, in the VUV region,
which is not directly detectable by the PMTs and needs to be
wavelength-shifted. The PMTs are coated with TPB
(tetraphenylbutadiene) that shifts the wavelength from the VUV to
the visible part of the spectrum (420 nm).  Most of the direct
scintillation light hits the sides of the detector rather than the
PMTs, therefore to enhance the light yield the sides of the
detector are covered with TPB coated reflectors. The charge
readout system is positioned above the liquid surface, in the
argon vapor. This system extracts the ionization charge from the
liquid phase into the vapor phase, provides charge multiplication
and finally detects the induced signals on a segmented anode.
Charge multiplication is necessary in order to produce a
detectable signal over noise for ionizing events with energies as
low as few keVs.  A multi-stage system of Large Electron
Multipliers (LEM) can provide the necessary thousandfold charge
multiplication. This LEM readout system is being developed on a smaller scale in dedicated setup (see section \ref{secLEM} ).
In order to drift free electrons in liquid argon over lengths of a
meter, the concentration of electronegative impurities in liquid
argon must be extremely small \cite{Icarus}
better than a part per billion (ppb). This level of
purity is obtained by recirculating the argon through a custom
made copper filter. For the complete description and design details of the ArDM detector see \cite{Rubbia1},\cite{Kaufamnn}.

\subsection{First test of ArDM with liquid Ar}
\begin{figure}[htp]
\centering
\includegraphics[angle=0,scale=0.49]{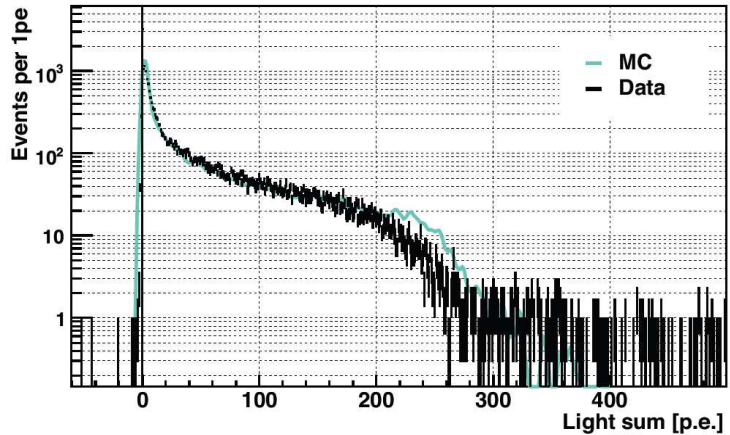}
\caption{\label{Fig.gammy}Energy spectrum from exposure to 511 keV gamma rays (MonteCarlo superimposed).}
\end{figure}

The ArDM detector was filled with liquid argon for the first time in May
2009. The run lasted for over a month. The light readout system was only partially completed and was deployed with 7 PMTs in place. The goals of this month long run were: test of the cryogenic system (cool down
and stable operation), achieve good and stable liquid argon purity,
commissioning and stable operation of the available light readout
system, provide a preliminary measurement of the light yield,
develop data reconstruction and provide data for benchmarking the
Monte Carlo simulation. Measurements of the light collection
efficiency were performed with two gamma sources,
$^{137}\textrm{Cs}$ and $ ^{22}\textrm{Na}$. Data were taken in a variety of trigger configurations. Figure
\ref{Fig.gammy} shows the energy spectrum in units of photoelectrons (p.e.) from the exposure to $^{22}$Na. The 511 keV gamma rays were externally tagged through detection
of the coincident 1274 keV and 511 keV gammas, the latter emitted back-to-back, with a NaI counter. From these measurements, a light yield of 0.5 p.e./keV was determined for this partially instrumented detector.
Offline it was determined that the rms noise is approximately 2
p.e. for the 7 PMTs, providing an energy threshold of few keVs.
For a preliminary  description of the measurements see also \cite{Regenfus}.

\section{Small setup for charge readout R\&D }\label{secLEM}
\begin{figure}[htp]
\includegraphics[angle=0,scale=0.35]{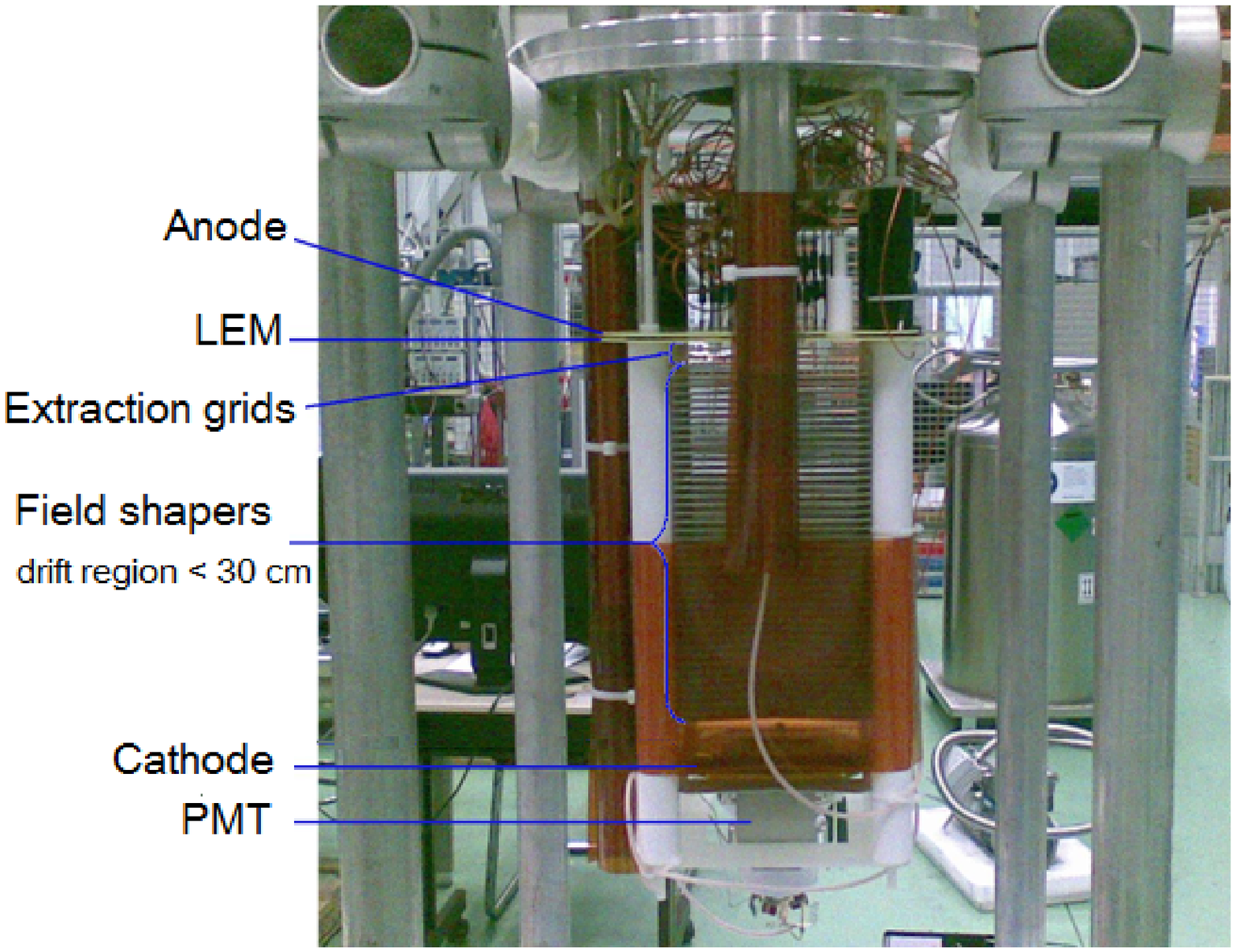}
\includegraphics[angle=0,scale=0.35]{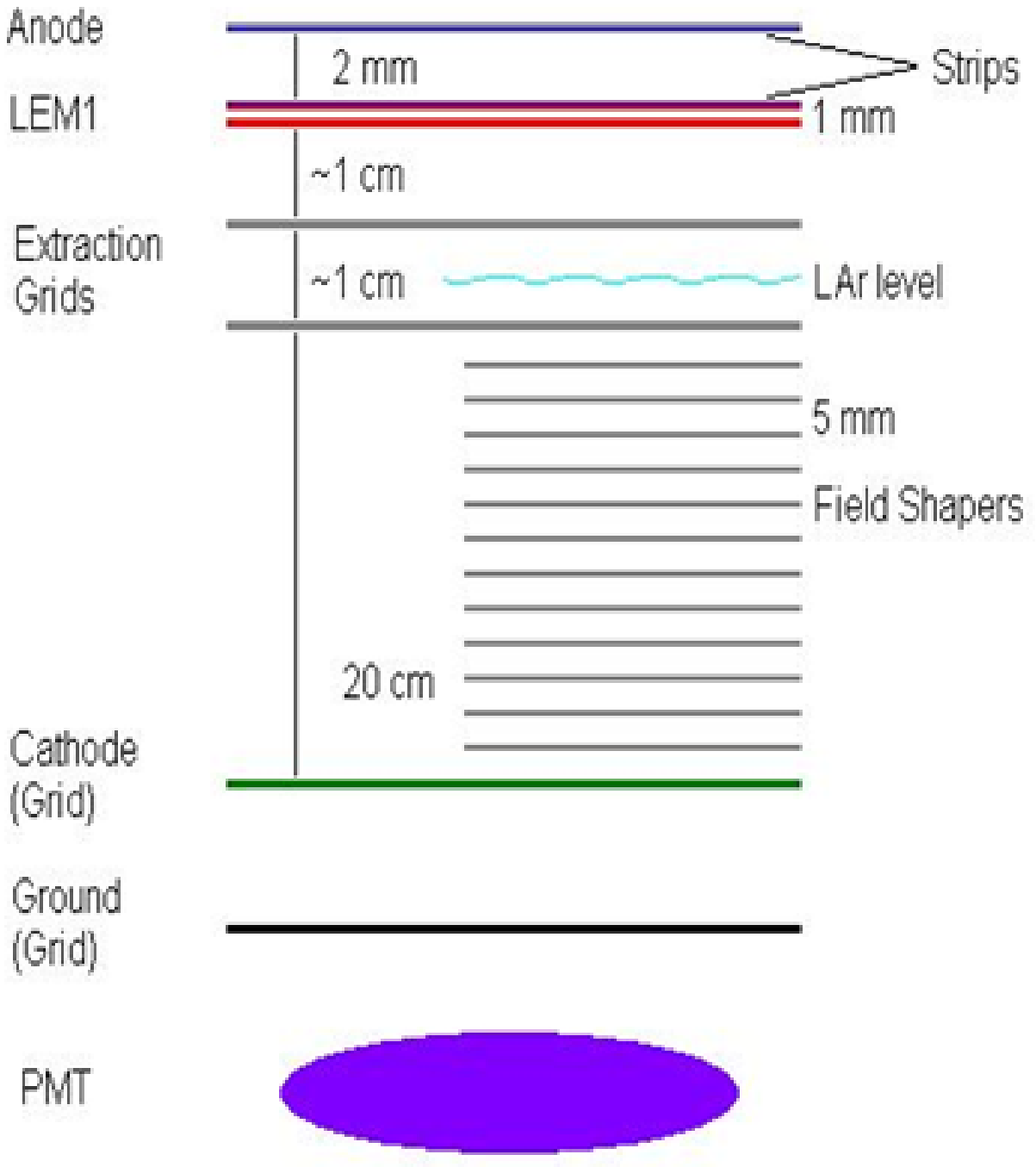}
\caption{\label{Fig.LEMfoto} Photograph (left) and schematics (right) of the argon LEM-TPC 3 lt setup.}
\end{figure}

\begin{figure}[htp]
\centering
\includegraphics[angle=0,scale=0.6]{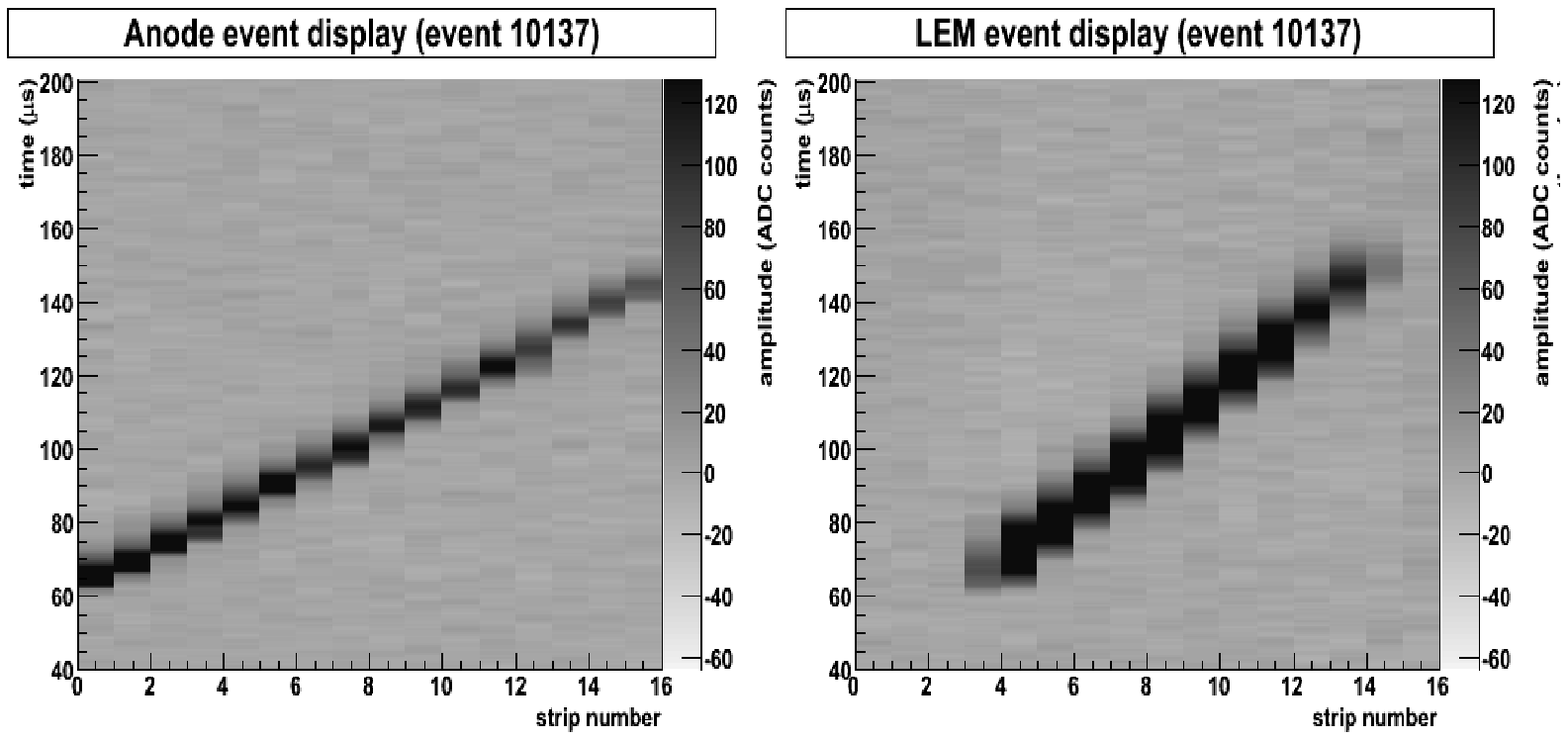}
\includegraphics[angle=0,scale=0.63]{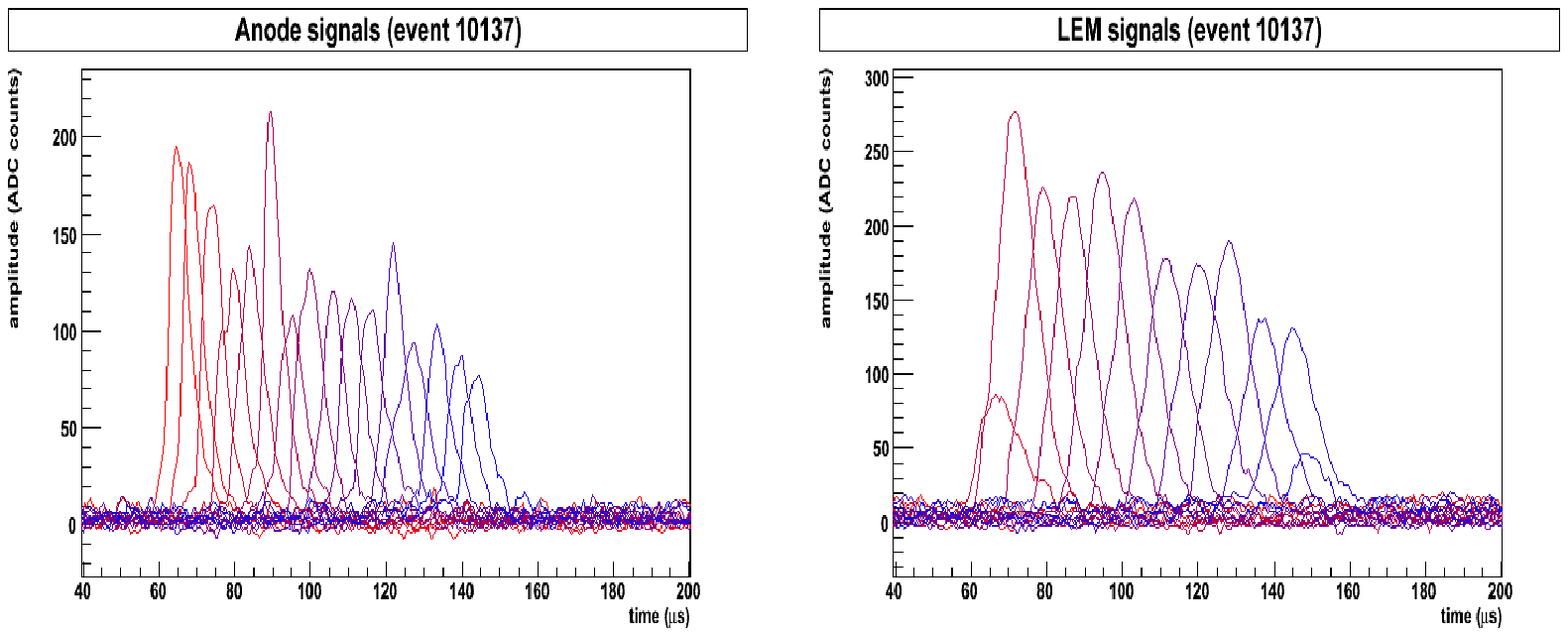}
\caption{\label{Fig.event} Typical cosmic muon event.}
\end{figure}

 A 3 l setup dedicated to R\&D work for the LEM
readout has been built and has been in operation for the past two
years \cite{Badertscher1 } \cite{Badertscher2}. It is a scaled
down version of the full ArDM detector, with a PMT installed below
the cathode, extraction grids at the liquid-vapor interface, LEM
for charge multiplication in the Ar vapor and a segmented anode
for readout. Charge multiplication happens in the LEM holes, where
there is a high enough electric field to start Townsend
multiplication \cite{Townsend} \cite{Sauli}. A picture and a sketch of the small LEM-TPC with one stage LEM device is shown in
Figure \ref{Fig.LEMfoto}. A LEM is a thick macroscopic
hole multiplier manufactured with standard PCB techniques. Several
different LEMs have been tested in this setup, differing for their
geometrical parameters, manufacturers, or both. Event shown here (Fig.\ref{Fig.event}) is for a one-stage LEM, 1 mm thick, double-sided
copper-clad plate drilled with a regular pattern of 500 $\mu$m
diameter holes, with a distance of 800 $\mu$m between the centers
of the neighboring holes. The LEM itself is segmented (strips
with 6 mm pitch), and after the LEM the charge is readout by a
segmented anode, with strips (6 mm pitch)  oriented orthogonally
with respect to LEM's strips, thus providing X-Y readout.

 This double phase argon LEM-TPC allows the reconstruction of the position of the ionizing event (the value of third
coordinate is assigned based on drift time). As an example, Figure
\ref{Fig.event} shows a typical cosmic muon track registered
during double phase operation.

\section{Conclusions and outlook}
The ArDM detector has had a first run in Spring 2009, which
allowed to test and commission the cryogenic system and a
partially completed light readout system. More tests are foreseen
for the remaining of year 2010, with the fully completed light
readout system in place. The detector will be operated as a full
double phase LAr TPC, with a direct readout of the ionization charge.
After completion of the R\&D  and commissioning phase, ArDM will
be moved to an underground, low background location to start the
Dark Matter science phase. Several underground laboratories are
under consideration, among them the SUNLAB laboratory at the
Sieroszowice-Polkowice mine in Poland.

%

\end{document}